\documentclass[aps,epsfig]{revtex4}
\usepackage{graphicx}
\usepackage{epstopdf}
\usepackage{latexsym}
\usepackage{amssymb}
\usepackage{textcomp}
\usepackage{amsmath}
\usepackage[center]{subfigure}
\usepackage{feynmp}
\begin{document}
\renewcommand\thesection{\arabic{section}}
\renewcommand\thesubsection{\thesection.\arabic{subsection}}
\renewcommand\thesubsubsection{\thesubsection.\arabic{subsubsection}}
 \newcommand{\bq}{\begin{equation}}
 \newcommand{\eq}{\end{equation}}
 \newcommand{\bqn}{\begin{eqnarray}}
 \newcommand{\eqn}{\end{eqnarray}}
 \newcommand{\nb}{\nonumber}
 \newcommand{\lb}{\label}
\title{A Lorentz Gauge Theory Of Gravity}
\author{Ahmad Borzou}
\email{ahmad_borzou@baylor.edu}
\unitlength 0.5mm
\affiliation{Experimental High Energy Group, Physics Department, Baylor
University, Waco, TX 76798-7316, USA}

\date{\today}

\begin{abstract}
We present a Lorentz gauge theory of gravity in which the metric is not dynamical. Spherically symmetric weak field solutions are
studied. We show that this solution contains the Schwarzschild spacetime at least to the first order of perturbation. Next, we present a special case of the theory where the Schwarzschild metric is an exact solution. It is also shown that the de Sitter space is an exact vacuum solution of this special case and as a result the theory is able to explain the expansion of the universe with no need for a dark energy. Within this special case, quantization of the theory is also studied, the basic Feynman diagrams are derived and renormalizability of the theory is studied using the power-counting method. We show that under a certain condition the theory is power-counting renormalizable. 
 
\end{abstract}

\maketitle

\section{Introduction}
Gauge theories have proven to be very successful in describing the fundamental interactions in physics. These can be
categorized into two different disciplines which work extremely well in terms of explaining the observations. On the one hand, 
the Standard model is a gauge theory of the group 
$SU(3)\times SU(2) \times U(1)$ which describes three of the physics interactions in terms of the geometry of internal spaces 
over spacetime. On the other hand, general relativity is a gauge theory of the Poincare group. Despite their similarity in 
being gauge theories, there is a glaring difference in their dynamical variables. In the former, the connections known as the 
vector bosons are the dynamical variables, while in the latter it is the metric and not the connections that is dynamical. 
Consequently, the Standard model Lagrangian is only a fourth order polynomial while that of general relativity
is not even a polynomial. One, however, can always expand the metric around a classical background which results in a
polynomial of infinite orders and the theory becomes more and more divergent as one goes to the higher orders in the
perturbative expansion.
This is why the Standard model has been successfully quantized while general relativity has not until now. An excellent review of 
the subject is provided in \cite{Weinberg}. On the basis of the Standard model achievements, a thorough investigation of the
relationship between the Standard model and gravitational theories might unveil important tips for the quantization of gravity. 
In this regard, people have scrutinized two main directions of research. The first direction is to find a duality between 
a gauge theory and gravity such as AdS/CFT correspondence introduced by Maldacena and further elaborated by others 
\cite{Maldacena,Gubser,Witten}. The present paper, however, lies within the second class, namely, attempts to
reformulate gravitational theory as a gauge theory. This is an approach to a gravitational theory that makes the gravitational
interactions look more like the interactions that are familiar from the Standard model of particle physics. This means
recasting the theory from a metric formulation to a formulation of connections of some internal spaces over spacetime. This avenue of
investigation began with the work of Utiyama \cite{Utiyama}. He localized the six parameters of the homogeneous Lorentz group
and showed that this consistently gives rise to the Einstein's general relativity. The idea was further extended by the work
of Sciama and Kibble \cite{Sciama, Kibble} by showing that a localized inhomogeneous Lorentz group realizes a well defined 
framework for gravity with torsion. There is a vast literature on the subject, acknowledging which would be an exhausting task. 
Here we only refer to two of the review papers \cite{Hehl, Ivanenko}. 
Although there has been an enormous progress in placing gravity and the Standard model onto one single footing, there are yet
some remaining differences. In doing so the main hurdle is the dynamical role of the metric. In the present paper we would like
to study a formulation of gravity in which metric is nondynamical. From the equivalence principle we know that at any point in
spacetime there is a free falling frame which comes with a unique feature, namely, being both a Lorentz and a 
coordinate frame. This fact enables one to split a given tetrad field into two parts. First, the part which contains the
angle between the free falling frame and the arbitrarily chosen Lorentz frame at that point. Second, the  part that contains 
the angle between the free falling frame and that associated with the arbitrarily chosen coordinates at that point. In a 
more rigorous language
\bq
e_{i \mu}=\eta^{\bar{k} \bar{l}} e_{i \bar{k}}e_{\bar{l}\mu},\nb
\eq
 where the bar indicates the free falling frame while the Latin index refers to the Lorentz frame and the Greek one refers to
 the coordinate system. Dynamics in the tetrad can be originated from either of the two constituents. Namely,

\bqn
\begin{cases}
\delta e_{i \mu}=\eta^{\bar{k} \bar{l}}e_{i \bar{k}}\delta e_{\bar{l} \mu} &  \text{Case I},\\
\delta e_{i \mu}=\eta^{\bar{k} \bar{l}}\delta e_{i \bar{k}}e_{\bar{l} \mu} &  \text{Case II}.\nb
\end{cases}
\eqn

The first case leads to the general theory of relativity and is not the subject of the present paper. The second case develops
no dynamics
in the metric. This is because the metric is independent of the choice of the Lorentz frame 
\bq
g_{\mu \nu}=\eta^{ij}e_{i \mu}e_{j \nu}
=\eta^{\bar{i}\bar{j}}e_{\bar{i} \mu}e_{\bar{j} \nu}.\nb
\eq
Therefore, $\delta g_{\mu \nu}=0$ in the latter case. This, however, doesn't mean that the metric is not affected at all.
As will be 
shown later, this approach establishes a formulation with a propagating spin connection. In the presence of a nonzero connection the 
difference between two neighboring free falling frames does not vanish and therefore spacetime departs from a Minkowskian form.

The present paper is organized as follows. A brief review of the tetrad formalism is presented in section 2. In section 3 
the Lorentz gauge theory is introduced by a Lagrangian, where the conservation laws as well as the field equations are derived. Here, like the very original work of Utiyama, we employ the spin connections, gravitational gauge fields, to preserve the local homogeneous Lorentz invariance. Next, a weak field solution is found for a spherically symmetric spacetime, where we show that it is the Schwarzschild solution at least to the first order of perturbation.  
In section 4, to make the theory more like the Standard model, a special case is introduced, where it is shown that the Schwarzschild as well as the de Sitter spaces are two exact vacuum solutions. Quantization of the theory is briefly studied next.
Propagator of the gauge field and also the principal vertices are derived as well. Then it is shown that 
under a certain condition, the theory is at least power-counting renormalizable. A conclusion is drawn at the end in section 5. 

\section{A brief review of the tetrad formalism}
General relativity successfully describes gravity in the macroscopic level. To this level matter is sufficiently well represented
by the energy momentum tensor. However, if one wishes to go down to the microscopic level, classical matter must be replaced 
by the elementary particles which are characterized not only by their masses but also by their spins. It is well understood 
that these elementary particles are explained by the Dirac Lagrangian. Therefore, one needs to deal with the Dirac matrices 
and spinors in a curved spacetime. This requires a generalization of their fundamental properties in the flat spacetime to more general forms that
hold in any curved spacetime. A simple breakthrough is to define a tangent space at any point on the manifold
and solve physics in those flat Lorentz spaces. It is now necessary to find a connection between the coordinate space and the 
flat Lorentz spaces. This goal is reached by introducing at each point of the manifold a set of four vector fields, called
tetrads. Now the Dirac Lagrangian reads
\bq
\lb{LDirac}
{\cal{L}}_{Dirac}=i\bar{\psi}\gamma^ie_i^{~\mu}\partial_{\mu}\psi-m\bar{\psi}\psi,
\eq
with $e_i^{~\mu}$ being the tetrad field.
Here the Latin indices indicate the Lorentz vectors while the Greek indices denote the covariant components of the Lorentz 
vectors, $\hat{e}_i$, in the curved spacetime. Both indices run from zero to three. \\
This Lagrangian is invariant under the global homogeneous Lorentz transformations. Under a local transformation the partial
derivative should be replaced by the following relation

\bq
\partial_{\mu} \rightarrow D_{\mu}=\partial_{\mu} - \frac{1}{2}S^{mn}A_{mn \mu},
\eq
where the commutator of the Dirac matrices, $S^{mn}=\frac{1}{4}[\gamma^m,\gamma^n]$, is the generator of the homogeneous
Lorentz group and the spin connection, $A_{mn \mu}$, is the gauge
preserving field. It is not hard to show that under homogeneous Lorentz transformations
\bq
\lb{deltaA}
\delta A_{mn \mu}=D_{\mu}\omega_{mn}=\partial_{\mu}\omega_{mn}-A_{mk\mu}\omega^k_{~n}-A_{nk\mu}\omega_m^{~~k}.
\eq
Here $\omega_{mn}$ is an antisymmetric tensor that can take any arbitrary value.
Since the Lorentz space is flat, the metric on the Lorentz space is always Minkowskian, with a zero covariant derivative 
in order to preserve angles. This makes the gauge field antisymmetric in the Lorentz indices. The equivalence of the connections
in the coordinate and the Lorentz spaces implies the tetrad postulate, which denotes that the covariant derivative of the tetrad field
is zero,
\bq
\lb{TetPos}
D_{\mu}e_{i \nu}=\partial_{\mu}e_{i \nu}-\Gamma^{\alpha}_{\mu \nu}e_{i \alpha}-A_{ij \mu}e^j_{~\nu}
=\nabla_{\mu}e_{i \nu}-A_{ij \mu}e^j_{~\nu}=0,
\eq
where $\Gamma^{\alpha}_{\mu \nu}$ are the metric compatible Christoffel symbols
\bq
\lb{Chris}
\Gamma^{\alpha}_{\mu \nu}=\frac{1}{2}g^{\alpha \beta}(\partial_{\nu}g_{\mu \beta}+\partial_{\mu}g_{\nu \beta}
-\partial_{\beta}g_{\mu\nu}).
\eq
In the present paper we solely work 
with a torsion free space indicating that the symbols are symmetric with respect to the two lower indices.
Using (\ref{TetPos}) the spin connections are
\bq
\lb{SpinCon}
A_{ij \mu}=e_j^{~\nu}\partial_{\mu}e_{i \nu}-\Gamma^{\alpha}_{\mu \nu}e_{i \alpha}e_j^{~\nu}.
\eq
Using the principle of equivalence we can define at each point $X$ an inertial coordinate system $\zeta^i$ in which equation of 
motion of a freely falling particle is 
\bq
\frac{d^2\zeta^i}{d\tau ^2}=0.
\eq
A straightforward calculation gives an equation for the Christoffel symbols
\bq
\Gamma^{\lambda}_{\mu \nu}= \frac{\partial x^{\lambda}}{\partial \zeta^{\alpha}}\frac{\partial ^2 \zeta^{\alpha}}
{\partial x_{\mu}\partial x_{\nu}}.
\eq
This can be used to find the locally inertial coordinates
\bq
\lb{xi}
\zeta^{i}(x)=e^i_{~\mu}(X)(x^{\mu}-X^{\mu})+e^i_{~\mu}(X)\Gamma ^{\mu}_{\alpha \beta}(x^{\alpha}-X^{\alpha})(x^{\beta}-X^{\beta})
+... .
\eq
More details can be found in \cite{Carroll,Weinberg2,Clarke}. 

\section{Homogeneous Lorentz Gauge Theory Of Gravity}
We formally define the homogeneous Lorentz gauge theory by the following action
\bqn
\lb{action}
S&=&\int e d^4x\Big[{\cal{L}}_{M}+{\cal{L}}_{A}\Big].
\eqn
Here $e$ is the determinant of the tetrad field while ${\cal{L}}_{M}$ specifies the interaction between matter and gravity and
is assumed to be the Dirac Lagrangian. 
A Lagrangian, ${\cal{L}}_{A}$, is needed as well to describe the gauge field itself. The action must remain invariant under both 
general coordinate and local homogeneous Lorentz transformations which in turn implies the conservation laws. Under an infinitesimal homogeneous Lorentz transformation 
\bq
\lb{LorentzTrans}
\delta S=\int d^4x\Big[\frac{\delta(e{\cal{L}}_{M})}{\delta \psi}\delta \psi
+\frac{\delta(e{\cal{L}}_{M})}{\delta A_{mn \mu}}\delta A_{mn \mu}
+\frac{\delta(e{\cal{L}}_{M})}{\delta e_{i \mu}}\delta e_{i \mu} \Big]=0.
\eq
The first term is the Dirac field equation and is zero. Using equation (\ref{deltaA}) the second term reads
\bq
\frac{\delta(e{\cal{L}}_{M})}{\delta A_{mn \mu}}\delta A_{mn \mu}=-D_{\mu}\Big(\frac{\delta(e{\cal{L}}_{M})}{\delta A_{mn \mu}} \Big)\omega_{mn},
\eq
where the surface term is neglected.
We also know that in the third term
\bq
\delta e_{i \mu}=\omega_{ij}e^j_{~\mu},
\eq
which is because the tetrad transforms like a vector under Lorentz transformations.
Therefore, equation (\ref{LorentzTrans}) reads
\bq
\lb{LorentzTrans2}
\delta S=-\int d^4x\Big[D_{ \mu}\Big(\frac{\delta(e{\cal{L}}_{M})}{\delta A_{mn \mu}}\Big)
-\frac{1}{2}\frac{\delta(e{\cal{L}}_{M})}{\delta e_{m \mu}}e^n_{~\mu}
+\frac{1}{2}\frac{\delta(e{\cal{L}}_{M})}{\delta e_{n \mu}}e^m_{~\mu} \Big]\omega_{mn}=0.
\eq
On the other hand $\omega_{mn}$ can take any arbitrary value implying that the bracket contains a zero. These altogether grant the conservation law of angular momentum
\bq
\lb{conservationLaw}
D_{ \mu}\Big(\frac{\delta(e{\cal{L}}_{M})}{\delta A_{mn \mu}}\Big)
-\frac{1}{2}\frac{\delta(e{\cal{L}}_{M})}{\delta e_{m \mu}}e^n_{~\mu}
+\frac{1}{2}\frac{\delta(e{\cal{L}}_{M})}{\delta e_{n \mu}}e^m_{~\mu}=0.
\eq

Before proceeding further and deriving the field
equations, the tetrad field should be investigated a little bit more. Because of the equivalence principle 
it is always possible to split a given tetrad field at any point $X$ into two parts
\bq
e_{i \mu}(X)=\eta^{\bar{j} \bar{k}}e_{i \bar{j}}(X)e_{\bar{k} \mu}(X).
\eq
This is because it is guaranteed that there exist a free falling frame whose coordinate system is locally Minkowskian, and as a result coincides with one of the possible Lorentz frames at that point which is what is shown with a bar in the equation
above and corresponds with a set of four orthogonal unit vectors, $\hat{e}_{\bar{i}}$. Components of these vectors in any arbitrary 
Lorentz frame are shown with $e_{i\bar{j~}}$. On the other hand, components of these free falling unit vectors in any 
arbitrary coordinate system is shown with $e_{\bar{k} \mu}$. 
An infinitesimal change in the tetrad field can be established in two ways. The first which is the subject of the present 
study is 
\bq
\lb{tetradvar}
\delta e_{i \mu}(X)=\eta^{\bar{j} \bar{k}}\delta e_{i \bar{j~}}(X)e_{\bar{k} \mu}(X).
\eq
The second one, which results in the theory of general relativity, is well investigated before 
\bq
\lb{caseII}
\delta e_{i \mu}(X)=\eta^{\bar{j} \bar{k}} e_{i \bar{j~}}(X)\delta e_{\bar{k} \mu}(X).
\eq

One of the consequences of equation (\ref{tetradvar}) is that $\delta g_{\mu \nu}=0$. This is because
$g_{\mu \nu}=\eta^{ij}e_{i \mu}e_{j \nu}=\eta^{\bar{i}\bar{j}}e_{\bar{i} \mu}e_{\bar{j} \nu}$ is independent of the chosen 
Lorentz frame.
Another consequence is that
\bq
\lb{delA}
\delta A_{ij \mu}=D_{\mu}(e_j^{~\nu}\delta e_{i \nu}).
\eq 
This is reached by varying (\ref{TetPos}) with respect to the tetrad
\bqn
&&\partial_{\mu}\delta e_{i \nu}-\delta\Gamma^{\alpha}_{\mu \nu}e_{i \alpha}-\Gamma^{\alpha}_{\mu \nu}\delta e_{i \alpha}-\delta A_{ij \mu}e^j_{~\nu}-A_{ij \mu}\delta e^j_{~\nu}=\nb\\
&&D_{\mu}\delta e_{i \nu}-\delta\Gamma^{\alpha}_{\mu \nu}e_{i \alpha}-\delta A_{ij \mu}e^j_{~\nu}=0,
\eqn
and the fact that
\bqn
&&\delta g_{\mu\nu}=0,\nb\\
&&\delta \Gamma^{\alpha}_{\beta \gamma}=0.
\eqn  
This equation can be used to show that the tetrad
field is not propagating at all. This is because a variation of the action (\ref{action}) with respect to the tetrad field reads
\bq
\lb{tetradfieldeq}
\frac{\delta(e {\cal{L}}_A)}{\delta e_{i\mu}}=
- \frac{\delta(e {\cal{L}}_M)}{\delta e_{i\mu}},
\eq
where 
\bqn
\delta(e {\cal{L}}_M)=\frac{\delta(e {\cal{L}}_M)}{\delta A_{ij\mu}}\delta A_{ij\mu}+\frac{\delta(e {\cal{L}}_M)}{\delta e_{i\mu}}\delta e_{i\mu}.
\eqn
Using equation (\ref{delA}) and neglecting the surface terms
\bqn
\delta(e {\cal{L}}_M)&=&-D_{\mu}\frac{\delta(e {\cal{L}}_M)}{\delta A_{ij\mu}}e_j^{~\nu}\delta e_{i \nu}+\frac{\delta(e {\cal{L}}_M)}{\delta e_{i\mu}}\delta e_{i\mu}\nb\\
&=&-D_{\mu}\frac{\delta(e {\cal{L}}_M)}{\delta A_{ij\mu}}e_j^{~\nu}\delta e_{i \nu}+\frac{\delta(e {\cal{L}}_M)}{\delta e_{i\mu}}(\frac{1}{2}e^j_{~\mu}e_j^{~\nu}\delta e_{i\nu}-\frac{1}{2}e^j_{~\mu}e_i^{~\nu}\delta e_{j\nu})\nb\\
&=&-\Big[D_{\mu}\frac{\delta(e {\cal{L}}_M)}{\delta A_{ij\mu}}-\frac{1}{2}\frac{\delta(e {\cal{L}}_M)}{\delta e_{i\mu}}e^j_{~\mu}+\frac{1}{2}\frac{\delta(e {\cal{L}}_M)}{\delta e_{j\mu}}e^i_{~\mu}\Big]e_j^{~\nu}\delta e_{i \nu}\nb\\
&=&0,
\eqn
where we have used 
\bqn
&&\delta e_{i\mu}=g^{\nu}_{\mu}\delta e_{i\nu}=e^j_{~\mu}e_j^{~\nu}\delta e_{i\nu},\nb\\
&&e_j^{~\nu}\delta e_{i \nu}=-e_i^{~\nu}\delta e_{j \nu},\nb\\
&&\delta\eta_{ij}=0,
\eqn
together with equation (\ref{conservationLaw}).
Therefore the right hand side, the source term, of equation (\ref{tetradfieldeq}) is zero which means no source exists to generate the tetrad field. We would like to 
emphasize that this result holds only if the variation path is given by equation (\ref{tetradvar}). 
If on the other hand the variation path is the one introduced by equation (\ref{caseII}), it results in the general
theory of relativity, which is well investigated. As is shown above, however, equation (\ref{tetradvar}) results in no propagation
of the tetrad field. Hence, in order to have a set of field equations, we are left with one option, namely, varying the action with respect to the spin connection and eliminating
the tetrad in terms of that. The difficulty now is to write $\delta e_{i \mu}$ in 
terms of $\delta A_{ij \mu}$. This problem can be solved by the use of the Lagrange multiplier method 
by inserting the tetrad postulate in the action as a constraint
\bq
\lb{LC}
{\cal{L}}_{C}=S^{\mu \nu i}D_{\mu}e_{i \nu},
\eq
where $S^{\mu \nu i}$ is the multiplier.  
Assuming conservation of parity, the most general Lagrangian for the gauge field is \cite{Hayashi, Daemi}
\bqn
\lb{LA}
{\cal{L}}_{A}&=&-\frac{1}{4}\Big(c_1 F_{\mu\nu ij}e^{i \mu}e^{j \nu} + c_2 F_{\mu\nu ij}F^{\mu \sigma ik}e^{j \nu}e_{k \sigma}
+c_3 F_{\sigma \nu mj}F_{\mu \alpha in}e^{j \nu}e^{i \mu}e^{m \sigma}e^{n \alpha}\nb \\
&&~~~+c_4F_{\mu \nu ij}F^{\alpha \beta mn}e^{i \mu}e^j_{~\beta}e_{m \alpha}e_n^{~\nu}+c_5F_{\mu \nu ij}F^{\mu\nu ij}\Big),
\eqn
where
\bq
\lb{strength}
F_{\mu\nu ij}=\partial_{\nu}A_{ij\mu}-\partial_{\mu}A_{ij\nu}+A_{i~~\mu}^{~m}A_{mj\nu}-A_{i~~\nu}^{~m}A_{mj\mu}.
\eq
In the following we assume $c_1=0$ since it involves an odd number of derivatives and leads to a non-propagating interaction. 
Field equations can be derived by varying the Lagrangians (\ref{LDirac}), (\ref{LC}), and (\ref{LA}) with respect to 
$e_{i \mu}$, $A_{ij \mu}$ and $S^{\mu \nu i}$. Variation with respect to $S^{\mu \nu i}$ returns the tetrad postulate.
Variation with respect to the gauge field reads
\bqn
\lb{Field1}
\frac{\delta(e{\cal{L}}_\textbf{Total})}{\delta A_{ij \mu}}&=&
\frac{1}{4}D_{\nu}\Big(\{c_2F^{\mu \sigma i k}e^{j \nu}e_{k \sigma}+c_3F_{\sigma \alpha mn}e^{i \mu}e^{j \nu}e^{n \alpha}e^{m \sigma}
+c_4F^{\alpha \beta mn}e_{m \alpha}e^j_{~\beta}e^{i \mu}e_n^{~\nu}-(i \leftrightarrow j)\}-(\mu \leftrightarrow \nu)\Big)\nb\\
&&+c_5D_{\nu}F^{\mu \nu ij}+\frac{\delta{\cal{L}}_{M}}{\delta A_{ij \mu}}-\frac{1}{2}S^{\mu \nu i}e^j_{~\nu}+\frac{1}{2}S^{\mu \nu j}e^i_{~\nu}=0.
\eqn
Here $\frac{\delta{\cal{L}}_{M}}{\delta A_{ij \mu}}$ is the spin angular momentum of matter while the last two terms are of angular momentum type and acceptable only if defined locally. Variation with respect to
$e_{i \alpha}$ reads 
\bqn
\lb{Field2}
\frac{\delta{\cal{L}}_\textbf{Total}}{\delta e_{i \alpha}}&=&
\frac{\delta{\cal{L}}_\textbf{Matter}}{\delta e_{i \alpha}}-D_{\beta}S^{\beta \alpha i}\nb\\
&&-\frac{1}{2}c_2F^{\beta \alpha ~ i}_{~~~j}F_{\beta \lambda}^{~~jk}e_k^{~\lambda}
-c_3F^{\mu \lambda mj}F^{\alpha \nu in}e_{m \mu}e_{n \nu}e_{j \lambda}\nb\\
&&-\frac{1}{2}c_4F^{\alpha \mu ij}F^{\nu \beta mn}e_{m \nu}e_{j \beta}e_{n \mu}
-\frac{1}{2}c_4F^{\mu\nu ji}F^{\beta\alpha mn}e_{j \mu}e_{m \beta}e_{n \nu}=0.
\eqn
Note that we already set $c_1=0$ and also $S^{\mu \nu i}$ is a non-propagating field, i.e., is zero outside of matter. The solution to equation (\ref{Field2}), by neglecting the second order terms in $F$, is

\bqn
S^{\alpha \beta i}=T^{\alpha \beta}\xi^i,
\eqn
where $T^{\alpha \beta}$ is the energy momentum tensor and $\xi^i$ is defined as follows

\bq
\lb{xi2}
\xi^i(x)=
\begin{cases}
e^i_{~\alpha}(X)(x^{\alpha}-X^{\alpha}) & x < \delta, \\
0 & x \geq \delta,
\end{cases}
\eq
where $\delta$ is assumed to be very small and $X$ refers to a local point. 

\subsection{Static Spherically Symmetric Case: A Weak Field Approximation}
In this part we would like to find a static spherically symmetric solution. An approximate approach is sufficient for our 
purposes. We start with the following tetrad field
\bqn
\lb{weakSch}
e_{i \mu}=
\begin{pmatrix}
\sqrt{a(r)}&~&~&~\\
~&\sqrt{b(r)}&~&~\\
~&~&r&~\\
~&~&~&rsin(\theta)
\end{pmatrix},
\eqn
where
\bqn
a&=&1+\delta a,\nb\\
b&=&1+\delta b, 
\eqn
with $\delta a $ and $ \delta b \ll 1$.\\
Here the results to the first order of perturbation in $\delta a$ and $\delta b$ are desired, and therefore for the rest of the section, only the first order terms will be kept. The Christoffel symbols, $\Gamma^{\lambda}_{\mu\nu}$, can be easily calculated
using (\ref{Chris})
\begin{align}
\Gamma^1_{00}&=\frac{1}{2}\delta a', & \Gamma^2_{12}&=\frac{1}{r},& \Gamma^1_{22}&=-r(1-\delta b), \nb\\
\Gamma^0_{01}&=\frac{1}{2}\delta a', & \Gamma^3_{13}&=\frac{1}{r},& \Gamma^1_{33}&=-rsin^2(\theta)(1-\delta b), \nb\\
\Gamma^1_{11}&=\frac{1}{2}\delta b', & \Gamma^3_{23}&=\frac{cos(\theta)}{sin(\theta)},& \Gamma^2_{33}&=-sin(\theta)cos(\theta), \nb\\
\end{align}
where prime indicates derivative with respect to r. The spin connections, $A_{ij\mu}$, using (\ref{SpinCon}) are

\begin{align}
A_{100}&=\frac{1}{2}\delta a',&  A_{122}&=1-\frac{1}{2}\delta b,\nb\\
A_{133}&=(1-\frac{1}{2}\delta b)sin(\theta),& A_{233}&=cos(\theta),\nb\\
\end{align}

and the strength tensor, $F_{\mu\nu ij}$, using (\ref{strength}) reads
\begin{align}
F_{1010}&=-\frac{1}{2}\delta a'', & F_{0220}&=\frac{1}{2}\delta a', & F_{0330}&=\frac{1}{2}sin(\theta)\delta a',\nb\\
F_{1221}&=-\frac{1}{2}\delta b', & F_{1331}&=-\frac{1}{2}sin(\theta)\delta b', & F_{3232}&=sin(\theta)\delta b.
\end{align} 
Here, and also in the rest of the paper, only nonzero components are shown. Inserting everything into equation (\ref{Field1}) and neglecting terms of second orders in $\delta a$ and $\delta b$
\bqn
\lb{eqII}
\begin{cases}
(c_2+2c_3+c_4+2c_5)\Big(r^3\delta a'''+2r^2\delta a''-2r\delta a'\Big)
-(c_2+4c_3+c_4)\Big(r^2\delta b''-2\delta b\Big)=\\
2r^3\Big(S^{0 0 1}e^0_{~0}-S^{0 1 0}e^1_{~1}\Big), & ~\\
~ & ~ \\
r^2\delta b''-2\delta b=\frac{c_2+4c_3+c_4}{3c_2+8c_3+3c_4+4c_5}\Big(r^3\delta a'''+2r^2\delta a''-2r\delta a'\Big). & ~
\end{cases}
\eqn
These two can be used to write down one of the two final equations
\bqn
\lb{desired eq I}
&&r^2\delta a'''+2r\delta a''-2\delta a'
=2r^2\lambda^{-1}\Big(S^{0 0 1}e^0_{~0}-S^{0 1 0}e^1_{~1}\Big),
\eqn
where
\bq
\lambda=\frac{(c_2+2c_3+c_4+2c_5)(3c_2+8c_3+3c_4+4c_5)-(c_2+4c_3+c_4)^2}{3c_2+8c_3+3c_4+4c_5},
\eq
and is a constant. 
The right hand side of (\ref{desired eq I}) is zero for a vacuum case. Therefore, the most general solution
is 
\bq
\delta a'=\frac{\alpha_1}{r^2}+\alpha_2r.
\eq
This solution should go to zero at large distances, which implies that $\alpha_2=0$. The other constant can be determined by 
comparing with the Schwarzschild solution
\bq
\alpha_1=2GM.
\eq
Using (\ref{eqII}) and considering that the right hand side of (\ref{desired eq I}) is zero, the other equation is 
\bq
r^2\delta b''-2\delta b=0,
\eq
with the most general solution 
\bq
\delta b = \frac{\beta_1}{r}+\beta_2 r^2.
\eq
In order to have a proper behavior at infinity, $\beta_2=0$. The other constant is 
\bq
\beta_1=2GM,
\eq
which comes from comparison with the Schwarzschild solution.

\section{A Special Case}
On the one hand, in the Standard model of particle physics the field equations are of the following form

\bq
D_{\mu}F^{\mu \nu} = J^{\nu},
\eq
where F, the field strength, has no direct contribution to the source, J, i.e., J $\neq$ J(F). On the other hand, it is strongly desired to make our gravitational theory as close to the Standard model as possible. That means the source of our theory should not depend on the strength tensor. In the theory presented above, the source can be read from (\ref{Field1})
\bq
\lb{source}
J^{\mu ij}=\frac{\delta{\cal{L}}_{M}}{\delta A_{ji \mu}}+S^{\mu \nu [i}e^{j]}_{~\nu},
\eq
where anti-symmetrization is denoted by a pair of square brackets and $S^{\mu \nu j}$ is determined through equation (\ref{Field2}), from which it can be deduced that by setting $c_1$ through $c_4$ to zero, the direct contribution of the strength field to the source can be eliminated. Therefore, we are left with one single term in the gauge field Lagrangian, equation (\ref{LA}), which defines the special case  
\bqn
\lb{NLA}
{\cal{L}}_{A}&=&-\frac{1}{4}c_5F_{\mu \nu ij}F^{\mu\nu ij}.
\eqn
The field equations now read 
\bqn
&&\frac{\delta{\cal{L}}_\textbf{Matter}}{\delta e_{i \alpha}}-D_{\beta}S^{\beta \alpha i}=0,\nb\\
&&c_5D_{\nu}F^{\mu \nu ij}=J^{\mu ij},
\eqn
with $J$ given by (\ref{source}).
The first equation implies the exact solution, $S^{\mu \nu i}=T^{\mu \nu}\xi^i$, which can be used to eliminate $S^{\mu \nu i}$ 
in the source term and reduce the whole set to 
\bqn
\lb{SimpleField}
c_5D^{\nu}F_{\mu\nu ij}&=&J_{\mu ij}\nb\\
&=&\frac{\delta{\cal{L}}_{M}}{\delta A^{ji \mu}}+\frac{1}{2}T_{\mu j}\xi_i-\frac{1}{2}T_{\mu i}\xi_j.
\eqn

\subsection{Static Spherically Symmetric Case: An Exact Solution}
For any proposed theory of gravity, it is crucial to address the experimental tests that general relativity has already passed and most of these experiments are performed within the solar system which is a static spherically symmetric case and this makes the subject specifically important. See \cite{Will} for a thorough review of the subject. The Schwarzschild metric, the solution to a static spherically symmetric space in GR, has explained all the relevant experiments and, consequently, should be the solution of any theory of gravity at least to some higher than one orders of perturbation since the first order is not sufficient to explain all the existing observations.
Fortunately it is not hard to show that this metric is an exact solution to the special case we have presented in this section. We start with the following tetrad 

\bqn
\lb{Sch}
e_{i \mu}=
\begin{pmatrix}
\sqrt{a(r)}&~&~&~\\
~&\frac{1}{\sqrt{a(r)}}&~&~\\
~&~&r&~\\
~&~&~&rsin(\theta)
\end{pmatrix}.
\eqn
The Christoffel symbols, $\Gamma^{\lambda}_{\mu\nu}$, are
\begin{align}
\Gamma^1_{00}&=\frac{1}{2}aa', & \Gamma^2_{12}&=\frac{1}{r},& \Gamma^1_{22}&=-ra, \nb\\
\Gamma^0_{01}&=\frac{1}{2}\frac{a'}{a}, & \Gamma^3_{13}&=\frac{1}{r},& \Gamma^1_{33}&=-rsin^2(\theta)a, \nb\\
\Gamma^1_{11}&=-\frac{1}{2}\frac{a'}{a}, & \Gamma^3_{23}&=\frac{cos(\theta)}{sin(\theta)},& \Gamma^2_{33}&=-sin(\theta)cos(\theta), \nb\\
\end{align}
where prime indicates derivative with respect to r. The spin connections, $A_{ij\mu}$, are
\begin{align}
A_{100}&=\frac{1}{2}a', & A_{122}&=\sqrt{a},\nb\\
A_{133}&=\sqrt{a}sin(\theta), & A_{233}&=cos(\theta),\nb\\
\end{align}
and the strength tensor, $F_{\mu\nu ij}$, is
\begin{align}
F_{1010}&=-\frac{1}{2}a'', & F_{0220}&=\frac{1}{2}\sqrt{a} a', & F_{0330}&=\frac{1}{2}\sqrt{a}sin(\theta)a',\nb\\
F_{1221}&=\frac{1}{2}\frac{a'}{\sqrt{a}}, & F_{1331}&=\frac{1}{2}sin(\theta)\frac{a'}{\sqrt{a}}, & F_{3232}&=sin(\theta)(1-a).
\end{align}
Substituting everything into (\ref{SimpleField}) and assuming a vacuum case results in two equations
\bqn
a'''+\frac{2}{r}a''-\frac{2}{r^2}a'=0,\nb\\
a''-\frac{2}{r^2}a+\frac{2}{r^2}=0.
\eqn
It is now easy to show that
\bq
a(r)=1-\frac{2GM}{r},
\eq 
satisfies both of the equations, i.e., the Schwarzchild metric is an exact solution of this special case of the theory. 

\subsection{Homogeneous Isotropic Case: A Cosmological Solution }
Another important subject that any theory of gravity should somehow address is a homogeneous and isotropic space described by the Friedmann-Lemaître-Robertson-Walker metric, or equivalently, the following tetrad
\bqn
\lb{tetrad-cos}
e_{i \mu}=a(t)
\begin{pmatrix}
a(t)^{-1}&~&~&~\\
~&1&~&~\\
~&~&r&~\\
~&~&~&rsin(\theta)
\end{pmatrix}.
\eqn
The Christoffel symbols, $\Gamma^{\lambda}_{\mu\nu}$, are
\begin{align}
\Gamma^1_{01}&=\frac{\dot{a}}{a},& \Gamma^2_{02}&=\frac{\dot{a}}{a},& \Gamma^3_{03}&=\frac{\dot{a}}{a}, \nb\\
\Gamma^0_{11}&=a\dot{a}, & \Gamma^2_{12}&=\frac{1}{r},& \Gamma^3_{13}&=\frac{1}{r}, \nb\\
\Gamma^0_{22}&=r^2a\dot{a}, & \Gamma^1_{22}&=-r,& \Gamma^3_{23}&=\frac{cos(\theta)}{sin(\theta)}, \nb\\
\Gamma^0_{33}&=r^2sin^2(\theta)a\dot{a}, & \Gamma^1_{33}&=-rsin^2(\theta),& \Gamma^2_{33}&=-cos(\theta)sin(\theta), \nb\\
\end{align}
where dot indicates derivative with respect to time. The spin connections, $A_{ij\mu}$, are
\begin{align}
A_{101}&=\dot{a}, & A_{022}&=-r\dot{a}, &A_{033}&=-rsin(\theta)\dot{a},\nb\\
A_{122}&=1, &A_{233}&=cos(\theta), & A_{133}&=sin(\theta),\nb\\
\end{align}
and the strength tensor, $F_{\mu\nu ij}$, is
\begin{align}
F_{1010}&=\ddot{a}, & F_{0220}&=-r\ddot{a}, & F_{0330}&=-rsin(\theta)\ddot{a},\nb\\
F_{1221}&=-r\dot{a}^2, & F_{1331}&=-rsin(\theta)\dot{a}^2, & F_{3232}&=r^2sin(\theta)\dot{a}^2.
\end{align}
The experimental data was gathered in 1998 when two independent groups of cosmologists observed that the universe is expanding with a positive rate. Within the context of general relativity this observation is commonly explained by introducing the cosmological constant, an unknown form of energy with negative pressure. Here in this paper we would like to show that without the help of the cosmological constant, our theory is able to explain the observation. The problem will be dramatically simpler for a vacuum case where no matter exist at all. Indeed this is not an irrelevant assumption to make as the matter density in the current epoch of the universe is almost negligible.  
Substituting all the pieces into (\ref{SimpleField}) and assuming $J^{\mu ij}=0$ results in one single equation

\bq
\dddot{a}+\frac{\dot{a}}{a}\ddot{a}-2(\frac{\dot{a}}{a})^2\dot{a}=0.
\eq
It turns out that the solution to this equation is 
\bq
a(t)=e^{Ht},
\eq
where $H=\frac{\dot{a}}{a}$ is a constant. This is exactly the de Sitter space which also can be achieved in general relativity. The only difference is that in general relativity the cosmological constant is needed to achieve this solution while in the present theory the solution holds for a vacuum case.  

\subsection{Feynman Rules And Renormalizability Of The Lorentz Gauge Theory Of Gravity}
Here we start from (\ref{SimpleField}) where 

\bqn
&&T_{\mu i}=e_{j\mu}e_{i\alpha}\frac{\delta{\cal{L}}_{M}}{\delta e_{j \alpha}}=
e_{j\mu}e_{i\alpha}i\bar{\psi}\gamma^jg^{\alpha \beta}\{\partial_{\beta}\psi-\frac{1}{2}S^{mn}A_{mn\beta}\psi\},\nb\\
&&\frac{\delta{\cal{L}}_{M}}{\delta A^{ij \mu}}=-\frac{i}{2}e_{m\mu}\bar{\psi}\gamma^mS_{ij}\psi.
\eqn

To further simplify the calculations, a flat background will be chosen. This in turn means
$e_{i \mu}=\delta_{i \mu}$, $g_{\mu \nu}=\eta_{\mu \nu}$ and $\Gamma^{\gamma}_{\mu \nu}=0$. Since the constraint 
in equation (\ref{LC}) has been taken care of in (\ref{SimpleField}), we choose to quantize  
using this field equation. This is despite the fact that the path integral approach is proven to be very strong 
method when working with gauge theories. 
In our approach one needs to take care of the gauge freedom because otherwise the inverse of 
the propagator would be singular and also unitarity may be violated. Therefore, the Lorentz gauge will be adopted in the following
\bq
\partial^{\nu}A_{ij \nu}=0,
\eq
which, together with (\ref{SimpleField}), leads to the following field equations
\bqn
\lb{finalfield}
\partial^2A_{ij\mu}=c_5^{-1}\Big(-\frac{\delta{\cal{L}}_{M}}{\delta A^{ij \mu}}+\frac{1}{2}T_{\mu j}\xi_i
-\frac{1}{2}T_{\mu i}\xi_j\Big)
-F(A)_{ij\mu},
\eqn
where 
\bqn
F(A)_{ij\mu}&=&(\eta^a_i \eta^b_j - \eta^a_j \eta^b_i)\Big(2\eta^{mn}\eta^{\alpha \beta}\eta^{\gamma}_{\mu}
-\eta^{\alpha\gamma}\eta^{mn}\eta^{\beta}_{\mu}\Big)A_{mb\alpha}\partial_{\beta}A_{an\gamma}\nb\\
&+&\eta^{\alpha \gamma}\eta^{\beta}_{\mu}\Big(2\eta^m_j\eta^{bs}\eta^a_i\eta^{nr}+\eta^{mn}\eta^b_i\eta^{ar}\eta^s_j
-\eta^{mn}\eta^b_j\eta^{ar}\eta^s_i\Big)A_{mb\alpha}A_{an\gamma}A_{rs\beta}.
\eqn

It is now necessary to find the propagator of the gauge field, $D_{ij\mu,mn\nu}(y-x)$. It should be antisymmetric in the consecutive Lorentz indices because the gauge field also has the same property. Moreover, it should satisfy the followings
\bqn
&&D_{ij\mu,mn\nu}(y-x)=D_{mn\nu,ij\mu}(y-x),\nb\\
&&A_{mn\nu}(x)=-\int d^4y A^{ij\mu}(y)\partial^2D_{ij\mu,mn\nu}(y-x).
\eqn
Therefore the propagator has the form
\bq
\lb{propag}
D_{ij\mu,mn\nu}(x-y)=\frac{1}{2}\eta_{\mu\nu}\Big(\eta_{mi}\eta_{nj}-\eta_{mj}\eta_{ni}\Big)D(x-y),
\eq
with $D(x-y)$ satisfying 
\bq
\partial^2D(x-y)=-\delta^4(x-y),
\eq
where the solution is 
\bq
D(x-y)=\int \frac{d^4q}{(4\pi)^4}\frac{e^{-iq.(x-y)}}{q^2+i\varepsilon}.
\eq

In the field of particle physics we are usually interested in scattering problems. A particle in the distant past is moving
toward the scattering area and is described by a plane wave, $\epsilon_{ij\mu}e^{-ik_i.x}$, at the beginning. We would like to know the final
state in the far future. This information is 
stored in the transition amplitudes known as the S matrix
\bq
\lb{smatrix}
S_{fi}=\lim_{t \to \infty} <A_{\textbf{final}}(\vec{x},t)|A_{\textbf{initial}}(\vec{x},t)>.
\eq
Here $A_{\textbf{final}}$ can be replaced by a plane wave, $\epsilon_{ij\mu}e^{-ik_f.x}$, when time goes to infinity. 
On the other hand, $A_{\textbf{initial}}$ is a plane 
wave only in the distant past, $\epsilon_{ij\mu}e^{-ik_i.x}$, and develops to a somewhat more complicated in the future 
\bq
\lb{ASolution}
A_{ij\mu}(x)= \epsilon_{ij\mu}e^{-ik_i.x}+\int d^4y D_{ij\mu,mn\nu}(x-y)\Big(c_{5}^{-1}(
\frac{\delta{\cal{L}}_{M}}{\delta A_{mn \nu}}-\frac{1}{2}T^{\nu n}\xi^m+\frac{1}{2}T^{\nu m}\xi^n)
+F(A)^{mn\nu}\Big),
\eq
where the Green's function method is used. 
This is itself an integral equation, but if the interactions are weak enough, we can solve it perturbatively and keep 
as many terms as needed. Equations (\ref{smatrix}) and (\ref{ASolution}) can be used to derive any possible interaction to any
desired order. Deriving all the possible interactions is out of the scope of the current work.
We instead are interested in finding all the Feynman rules of the theory. These are the vertices and the propagator with which 
all the other interactions can be built and are also sufficient to investigate the renormalizability of the theory. 
The propagator is already derived and in the momentum space reads

\begin{fmffile}{propag}
\bqn  
  \parbox{20mm}{\begin{fmfgraph*}(40,40)
    \fmfleft{l}
    \fmf{wiggly,label=$A$}{l,r}
    
    \fmfright{r}
     \end{fmfgraph*}}
& = & \frac{1}{2}\frac{\eta_{\mu\nu}\Big(\eta_{mi}\eta_{nj}-\eta_{mj}\eta_{ni}\Big)}{q^2+i \varepsilon}.
\eqn
     \end{fmffile}
The self interactions are cyphered in $F(A)^{mn\nu}$. These are
\begin{fmffile}{firstvertex}
 \bqn 
 \lb{vert1}
 \parbox{20mm}{\begin{fmfgraph}(40,40)
  \fmfleft{i1,i2}
  \fmf{wiggly}{i1,v,o}
  \fmf{wiggly}{i2,v}
   \fmfright{o}
 \end{fmfgraph}}
 & = & iq_{_{\beta}}(\eta^a_i \eta^b_j - \eta^a_j \eta^b_i)\Big(2\eta^{mn}\eta^{\alpha \beta}\eta^{\gamma}_{\mu}
-\eta^{\alpha\gamma}\eta^{mn}\eta^{\beta}_{\mu}\Big),\nb\\
 \parbox{20mm}{\begin{fmfgraph}(40,40)
  \fmfleft{i1,i2}
  \fmf{wiggly}{i1,v,o1}
  \fmf{wiggly}{i2,v,o2}
   \fmfright{o1,o2}
 \end{fmfgraph}}
 & = & \eta^{\alpha \gamma}\eta^{\beta}_{\mu}\Big(2\eta^m_j\eta^{bs}\eta^a_i\eta^{nr}+\eta^{mn}\eta^b_i\eta^{ar}\eta^s_j
-\eta^{mn}\eta^b_j\eta^{ar}\eta^s_i\Big).
 \eqn
\end{fmffile}
Since all the spin connections that appear here have the same ranking, a permutation over them is in order. However, extra
care should be taken when field equations are used for the sake of quantization. One of the fields, with indices $(i,j,\mu)$,
is already distributed over all the legs of the diagrams. Therefore, only the remaining fields need to be permuted.
There are also two types of interactions with matter

\begin{fmffile}{spin}
 \bqn
 \lb{vert2}
 \parbox{20mm}{\begin{fmfgraph}(40,40)
  \fmfleft{i1,i2}
  \fmf{fermion}{i1,v,o}
  \fmf{wiggly}{i2,v}
   \fmfright{o}
 \end{fmfgraph}}
  & = &\frac{1}{2}c_5^{^{-1}}\Big(q_{_{\alpha}}\delta^{j\nu}\delta^{n\alpha}\gamma_j\xi^m 
  -q_{_{\alpha}}\delta^{j\nu}\delta^{m\alpha}\gamma_j\xi^n
  -i\delta^{k\nu}\gamma_kS^{mn}\Big),\nb\\
  \parbox{20mm}{\begin{fmfgraph}(40,40)
  \fmfleft{i1,i2}
  \fmf{fermion}{i1,v,o1}
  \fmf{wiggly}{i2,v,o2}
   \fmfright{o1,o2}
 \end{fmfgraph}}
 & = &\frac{i}{4}c_5^{^{-1}}\Big(\delta^{j \nu}\delta^{n \beta}\gamma_jS^{kl}\xi^m
 -\delta^{j \nu}\delta^{m \beta}\gamma_jS^{kl}\xi^n\Big).
 \eqn
\end{fmffile}
In order to preserve the gauge invariance in the presence of Feynman diagrams with loops, Faddeev-Popov ghost fields must
be introduced and utilized as well.  
At this point we can start our investigation into the renormalizability of the theory. A detailed study of the
subject is out of 
the scope of the present paper. We instead use the simple method of power-counting which only gives an idea about the
divergences and can't be used as an alternative to an exact proof. A good description of the subject is given in \citep{Weinberg3}.
In a given Feynman diagram of any order, there exist L number of loops, I number of internal lines, 
E number of external lines and V number of vertices. 
The superficial degree of divergence in four dimensions reads
\bq
\lb{diverg}
D=4L+\sum_i v_i(d_i-w_i)-I_f-2I_A.
\eq
Here summation is over the four vertices given by (\ref{vert1}) and (\ref{vert2}), and $v_i$ is the number of such vertices
in the diagram while $d_i$ is the number of derivatives in the $i$th vertex. 
Also, $w_i$ is zero for the vertices which contain no $\xi$, namely (\ref{vert1}), otherwise it is the momentum
dependence, if any, of $\xi$, i.e., $\xi \propto q^{-w}$. The source of this momentum dependence is not known at this point. One naive way to
achieve it, is to assume $\delta$ in equation (\ref{xi2}) is energy dependent.
The subscripts f and A indicate fermionic field and the gauge field respectively. 
It is now required to express the superficial degree of divergence in terms of the number of external lines and vertices. Here the following identities prove useful
\bqn
1&=&L+V-I,\nb\\
E_{(A/f)}&=&\sum_i n_{(A/f)}^{(i)}v_i-2I_{(A/f)},\nb\\
I&=&I_f+I_A,\nb\\
V&=&\sum_i v_i,
\eqn
where the subscript $(A/f)$ means either fermionic field or the gauge field and $n_{(A/f)}^{(i)}$ refers to the number of
fermionic
or gauge fields at the vertex labeled by i. Gathering all the pieces, the superficial degree of divergence can be rewritten as 
\bqn
D&=&4(1-\sum_i v_i +I_f+I_A)+\sum_i v_i(d_i-w_i)-I_f-2I_A\nb\\
&=&4+3I_f+2I_A-\sum_i v_i(4+w_i-d_i)\nb\\
&=&4+\frac{3}{2}(\sum_i v_i n_f^{(i)}-E_f)+\frac{2}{2}(\sum_i v_i n_A^{(i)}-E_A)-\sum_i v_i(4+w_i-d_i)\nb\\
&=&4-E_A-\frac{3}{2}E_f-\sum_i v_i(4-n_A^i-\frac{3}{2}n_f^i+w_i-d_i).
\eqn
In principle we can have a graph with as many number of vertices as wanted. In a renormalizable theory, the superficial degree
of divergence does not increase with the order in the perturbation theory. This increase does not happen in our case only if at any given vertex 
\bq
\lb{con}
4-n_A^i-\frac{3}{2}n_f^i+w_i-d_i \geqslant 0.
\eq
This factor is zero for both of vertices in (\ref{vert1}) since $(n_A, n_f, w, d)$ is (3, 0, 0, 1) in the first vertex and (4, 0, 0, 0) in the second one. On the other hand we have (1, 2, $w$, 1) for the first two terms in the first vertex in (\ref{vert2}) and (1,~2,~0,~0) for the last term and (2, 2, $w$, 0) for the second vertex. Hence, (\ref{con}) holds for the vertices in
(\ref{vert2}) only if $w_i \geqslant 1$. As mentioned above, this can be achieved if $\delta \propto \frac{1}{q^w}$ in equation (\ref{xi2}). 
Investigation of methods by which this momentum behavior can be reached is beyond the scope of this work and is left for future 
studies. Although the renormalizability of the theory has not been proved, under this condition, the power-counting method
suggests a good high energy behavior for the theory. 

\section{Conclusions}
In this paper we have presented a Lorentz gauge formulation of gravity in which the metric has no dynamics. To achieve this, we
have used the equivalence principle that assures the existence of a free falling frame whose coordinate system is locally
Minkowskian. Therefore, at any point in spacetime there always exist a frame which is both of coordinate and Lorentz types. This leads to the fact that a tetrad field can be split into two parts, namely, 
$e_{i \mu}=\eta^{\bar{k} \bar{l}} e_{i \bar{k}}e_{\bar{l}\mu}$, where the free falling frame has been 
indicated with a bar. A variation in the tetrad field can therefore stem from any of
the two constituents. One leads to the Einstein's theory of gravity while the other to a formulation with no dynamics for metric. 
Because of 
the spectacular success of the Standard model of particle physics both in terms of experiments and renormalizability, we have 
investigated the formulation that is more analogous to the Standard model, the latter case, within which we have shown that a variation of the action with
respect to the tetrad results in the trivial angular momentum conservation equation where there exist no source for the resulting field
equations. Consequently, the field equations have been derived by varying the action with respect to the spin connections where the Lagrange multiplier 
method has been used to impose the tetrad postulate and eliminate the tetrad as a function of the spin connection.

We have also investigated a spherically symmetric weak field solution and showed that to the first order of perturbation, it is
in agreement with the Schwarzschild solution. A special case of the theory is also presented where the Schwarzschild metric is an exact solution. Moreover, a homogeneous and isotropic space has also been studied within this special case. We have shown that there exist a natural exponentially expanding vacuum solution where cosmological constant or any other type of dark energy is absent. 
In addition, quantization of the theory has been studied briefly and all the basic Feynman diagrams have been derived. We also
have shown that the theory is power-counting renormalizable if a certain condition is met.

~\\{\bf Acknowledgements:}
We are grateful to professors Gerald Cleaver and Hamid Reza Sepangi for careful reading
of the draft and useful suggestions. Our special thanks are
due to Jay Dittmann and Kenichi Hatakeyama for their continued support and also not minding us spending enormous amount of time
on the present paper while
working under their supervision with the CMS collaboration at the LHC. We also wish to express our gratitude to professor Mohammad Mehdi Sheikh-Jabbari for many email exchanges on the subject and his useful suggestions.

\end{document}